\documentclass[sigconf, screen]{acmart}
\AtBeginDocument{%
  \providecommand\BibTeX{{%
    \normalfont B\kern-0.5em{\scshape i\kern-0.25em b}\kern-0.8em\TeX}}}

\setcopyright{rightsretained}
\copyrightyear{2024}
\acmYear{2024}

\acmConference[arXiv]{arXiv: arXiv Preprint}{2024}{Atlanta, GA}
\acmBooktitle{arXiv: arXiv Preprint, 2024, Atlanta, GA}

\usepackage[most]{tcolorbox}
\usepackage{mathtools}
\usepackage{wrapfig}
\usepackage{multicol}

\usepackage{totcount} %

\usepackage{subcaption}
\usepackage{enumitem}
\usepackage{rotating}
\usepackage{microtype}
\usepackage{bm}
\usepackage{soul}
\usepackage{wrapfig}

\usepackage[varqu]{zi4}
\usepackage{units}
\usepackage{booktabs}
\usepackage{colortbl}

\usepackage{supertabular}
\usepackage{rotating}
\usepackage{array}

\usepackage{eqnarray}
\usepackage{bbm}

\definecolor{redOV}{RGB}{255, 235, 238}
\definecolor{redI}{RGB}{255, 205, 210}
\definecolor{redII}{RGB}{239, 154, 154}
\definecolor{redIII}{RGB}{229, 115, 115}
\definecolor{redIV}{RGB}{239, 83, 80}
\definecolor{redV}{RGB}{244, 67, 54}
\definecolor{redVI}{RGB}{229, 57, 53}
\definecolor{redVII}{RGB}{211, 47, 47}
\definecolor{redVIII}{RGB}{198, 40, 40}
\definecolor{redIX}{RGB}{183, 28, 28}
\definecolor{redAI}{RGB}{255, 138, 128}
\definecolor{redAII}{RGB}{255, 82, 82}
\definecolor{redAIV}{RGB}{255, 23, 68}
\definecolor{redAVII}{RGB}{213, 0, 0}

\definecolor{pinkOV}{RGB}{252, 228, 236}
\definecolor{pinkI}{RGB}{248, 187, 208}
\definecolor{pinkII}{RGB}{244, 143, 177}
\definecolor{pinkIII}{RGB}{240, 98, 146}
\definecolor{pinkIV}{RGB}{236, 64, 122}
\definecolor{pinkV}{RGB}{233, 30, 99}
\definecolor{pinkVI}{RGB}{216, 27, 96}
\definecolor{pinkVII}{RGB}{194, 24, 91}
\definecolor{pinkVIII}{RGB}{173, 20, 87}
\definecolor{pinkIX}{RGB}{136, 14, 79}
\definecolor{pinkAI}{RGB}{255, 128, 171}
\definecolor{pinkAII}{RGB}{255, 64, 129}
\definecolor{pinkAIV}{RGB}{245, 0, 87}
\definecolor{pinkAVII}{RGB}{197, 17, 98}

\definecolor{purpleOV}{RGB}{243, 229, 245}
\definecolor{purpleI}{RGB}{225, 190, 231}
\definecolor{purpleII}{RGB}{206, 147, 216}
\definecolor{purpleIII}{RGB}{186, 104, 200}
\definecolor{purpleIV}{RGB}{171, 71, 188}
\definecolor{purpleV}{RGB}{156, 39, 176}
\definecolor{purpleVI}{RGB}{142, 36, 170}
\definecolor{purpleVII}{RGB}{123, 31, 162}
\definecolor{purpleVIII}{RGB}{106, 27, 154}
\definecolor{purpleIX}{RGB}{74, 20, 140}
\definecolor{purpleAI}{RGB}{234, 128, 252}
\definecolor{purpleAII}{RGB}{224, 64, 251}
\definecolor{purpleAIV}{RGB}{213, 0, 249}
\definecolor{purpleAVII}{RGB}{170, 0, 255}

\definecolor{deeppurpleOV}{RGB}{237, 231, 246}
\definecolor{deeppurpleI}{RGB}{209, 196, 233}
\definecolor{deeppurpleII}{RGB}{179, 157, 219}
\definecolor{deeppurpleIII}{RGB}{149, 117, 205}
\definecolor{deeppurpleIV}{RGB}{126, 87, 194}
\definecolor{deeppurpleV}{RGB}{103, 58, 183}
\definecolor{deeppurpleVI}{RGB}{94, 53, 177}
\definecolor{deeppurpleVII}{RGB}{81, 45, 168}
\definecolor{deeppurpleVIII}{RGB}{69, 39, 160}
\definecolor{deeppurpleIX}{RGB}{49, 27, 146}
\definecolor{deeppurpleAI}{RGB}{179, 136, 255}
\definecolor{deeppurpleAII}{RGB}{124, 77, 255}
\definecolor{deeppurpleAIV}{RGB}{101, 31, 255}
\definecolor{deeppurpleAVII}{RGB}{98, 0, 234}

\definecolor{indigoOV}{RGB}{232, 234, 246}
\definecolor{indigoI}{RGB}{197, 202, 233}
\definecolor{indigoII}{RGB}{159, 168, 218}
\definecolor{indigoIII}{RGB}{121, 134, 203}
\definecolor{indigoIV}{RGB}{92, 107, 192}
\definecolor{indigoV}{RGB}{63, 81, 181}
\definecolor{indigoVI}{RGB}{57, 73, 171}
\definecolor{indigoVII}{RGB}{48, 63, 159}
\definecolor{indigoVIII}{RGB}{40, 53, 147}
\definecolor{indigoIX}{RGB}{26, 35, 126}
\definecolor{indigoAI}{RGB}{140, 158, 255}
\definecolor{indigoAII}{RGB}{83, 109, 254}
\definecolor{indigoAIV}{RGB}{61, 90, 254}
\definecolor{indigoAVII}{RGB}{48, 79, 254}

\definecolor{blueOV}{RGB}{227, 242, 253}
\definecolor{blueI}{RGB}{187, 222, 251}
\definecolor{blueII}{RGB}{144, 202, 249}
\definecolor{blueIII}{RGB}{100, 181, 246}
\definecolor{blueIV}{RGB}{66, 165, 245}
\definecolor{blueV}{RGB}{33, 150, 243}
\definecolor{blueVI}{RGB}{30, 136, 229}
\definecolor{blueVII}{RGB}{25, 118, 210}
\definecolor{blueVIII}{RGB}{21, 101, 192}
\definecolor{blueIX}{RGB}{13, 71, 161}
\definecolor{blueAI}{RGB}{130, 177, 255}
\definecolor{blueAII}{RGB}{68, 138, 255}
\definecolor{blueAIV}{RGB}{41, 121, 255}
\definecolor{blueAVII}{RGB}{41, 98, 255}

\definecolor{lightblueOV}{RGB}{225, 245, 254}
\definecolor{lightblueI}{RGB}{179, 229, 252}
\definecolor{lightblueII}{RGB}{129, 212, 250}
\definecolor{lightblueIII}{RGB}{79, 195, 247}
\definecolor{lightblueIV}{RGB}{41, 182, 246}
\definecolor{lightblueV}{RGB}{3, 169, 244}
\definecolor{lightblueVI}{RGB}{3, 155, 229}
\definecolor{lightblueVII}{RGB}{2, 136, 209}
\definecolor{lightblueVIII}{RGB}{2, 119, 189}
\definecolor{lightblueIX}{RGB}{1, 87, 155}
\definecolor{lightblueAI}{RGB}{128, 216, 255}
\definecolor{lightblueAII}{RGB}{64, 196, 255}
\definecolor{lightblueAIV}{RGB}{0, 176, 255}
\definecolor{lightblueAVII}{RGB}{0, 145, 234}

\definecolor{cyanOV}{RGB}{224, 247, 250}
\definecolor{cyanI}{RGB}{178, 235, 242}
\definecolor{cyanII}{RGB}{128, 222, 234}
\definecolor{cyanIII}{RGB}{77, 208, 225}
\definecolor{cyanIV}{RGB}{38, 198, 218}
\definecolor{cyanV}{RGB}{0, 188, 212}
\definecolor{cyanVI}{RGB}{0, 172, 193}
\definecolor{cyanVII}{RGB}{0, 151, 167}
\definecolor{cyanVIII}{RGB}{0, 131, 143}
\definecolor{cyanIX}{RGB}{0, 96, 100}
\definecolor{cyanAI}{RGB}{132, 255, 255}
\definecolor{cyanAII}{RGB}{24, 255, 255}
\definecolor{cyanAIV}{RGB}{0, 229, 255}
\definecolor{cyanAVII}{RGB}{0, 184, 212}

\definecolor{tealOV}{RGB}{224, 242, 241}
\definecolor{tealI}{RGB}{178, 223, 219}
\definecolor{tealII}{RGB}{128, 203, 196}
\definecolor{tealIII}{RGB}{77, 182, 172}
\definecolor{tealIV}{RGB}{38, 166, 154}
\definecolor{tealV}{RGB}{0, 150, 136}
\definecolor{tealVI}{RGB}{0, 137, 123}
\definecolor{tealVII}{RGB}{0, 121, 107}
\definecolor{tealVIII}{RGB}{0, 105, 92}
\definecolor{tealIX}{RGB}{0, 77, 64}
\definecolor{tealAI}{RGB}{167, 255, 235}
\definecolor{tealAII}{RGB}{100, 255, 218}
\definecolor{tealAIV}{RGB}{29, 233, 182}
\definecolor{tealAVII}{RGB}{0, 191, 165}

\definecolor{greenOV}{RGB}{232, 245, 233}
\definecolor{greenI}{RGB}{200, 230, 201}
\definecolor{greenII}{RGB}{165, 214, 167}
\definecolor{greenIII}{RGB}{129, 199, 132}
\definecolor{greenIV}{RGB}{102, 187, 106}
\definecolor{greenV}{RGB}{76, 175, 80}
\definecolor{greenVI}{RGB}{67, 160, 71}
\definecolor{greenVII}{RGB}{56, 142, 60}
\definecolor{greenVIII}{RGB}{46, 125, 50}
\definecolor{greenIX}{RGB}{27, 94, 32}
\definecolor{greenAI}{RGB}{185, 246, 202}
\definecolor{greenAII}{RGB}{105, 240, 174}
\definecolor{greenAIV}{RGB}{0, 230, 118}
\definecolor{greenAVII}{RGB}{0, 200, 83}

\definecolor{lightgreenOV}{RGB}{241, 248, 233}
\definecolor{lightgreenI}{RGB}{220, 237, 200}
\definecolor{lightgreenII}{RGB}{197, 225, 165}
\definecolor{lightgreenIII}{RGB}{174, 213, 129}
\definecolor{lightgreenIV}{RGB}{156, 204, 101}
\definecolor{lightgreenV}{RGB}{139, 195, 74}
\definecolor{lightgreenVI}{RGB}{124, 179, 66}
\definecolor{lightgreenVII}{RGB}{104, 159, 56}
\definecolor{lightgreenVIII}{RGB}{85, 139, 47}
\definecolor{lightgreenIX}{RGB}{51, 105, 30}
\definecolor{lightgreenAI}{RGB}{204, 255, 144}
\definecolor{lightgreenAII}{RGB}{178, 255, 89}
\definecolor{lightgreenAIV}{RGB}{118, 255, 3}
\definecolor{lightgreenAVII}{RGB}{100, 221, 23}

\definecolor{limeOV}{RGB}{249, 251, 231}
\definecolor{limeI}{RGB}{240, 244, 195}
\definecolor{limeII}{RGB}{230, 238, 156}
\definecolor{limeIII}{RGB}{220, 231, 117}
\definecolor{limeIV}{RGB}{212, 225, 87}
\definecolor{limeV}{RGB}{205, 220, 57}
\definecolor{limeVI}{RGB}{192, 202, 51}
\definecolor{limeVII}{RGB}{175, 180, 43}
\definecolor{limeVIII}{RGB}{158, 157, 36}
\definecolor{limeIX}{RGB}{130, 119, 23}
\definecolor{limeAI}{RGB}{244, 255, 129}
\definecolor{limeAII}{RGB}{238, 255, 65}
\definecolor{limeAIV}{RGB}{198, 255, 0}
\definecolor{limeAVII}{RGB}{174, 234, 0}

\definecolor{yellowOV}{RGB}{255, 253, 231}
\definecolor{yellowI}{RGB}{255, 249, 196}
\definecolor{yellowII}{RGB}{255, 245, 157}
\definecolor{yellowIII}{RGB}{255, 241, 118}
\definecolor{yellowIV}{RGB}{255, 238, 88}
\definecolor{yellowV}{RGB}{255, 235, 59}
\definecolor{yellowVI}{RGB}{253, 216, 53}
\definecolor{yellowVII}{RGB}{251, 192, 45}
\definecolor{yellowVIII}{RGB}{249, 168, 37}
\definecolor{yellowIX}{RGB}{245, 127, 23}
\definecolor{yellowAI}{RGB}{255, 255, 141}
\definecolor{yellowAII}{RGB}{255, 255, 0}
\definecolor{yellowAIV}{RGB}{255, 234, 0}
\definecolor{yellowAVII}{RGB}{255, 214, 0}

\definecolor{amberOV}{RGB}{255, 248, 225}
\definecolor{amberI}{RGB}{255, 236, 179}
\definecolor{amberII}{RGB}{255, 224, 130}
\definecolor{amberIII}{RGB}{255, 213, 79}
\definecolor{amberIV}{RGB}{255, 202, 40}
\definecolor{amberV}{RGB}{255, 193, 7}
\definecolor{amberVI}{RGB}{255, 179, 0}
\definecolor{amberVII}{RGB}{255, 160, 0}
\definecolor{amberVIII}{RGB}{255, 143, 0}
\definecolor{amberIX}{RGB}{255, 111, 0}
\definecolor{amberAI}{RGB}{255, 229, 127}
\definecolor{amberAII}{RGB}{255, 215, 64}
\definecolor{amberAIV}{RGB}{255, 196, 0}
\definecolor{amberAVII}{RGB}{255, 171, 0}

\definecolor{orangeOV}{RGB}{255, 243, 224}
\definecolor{orangeI}{RGB}{255, 224, 178}
\definecolor{orangeII}{RGB}{255, 204, 128}
\definecolor{orangeIII}{RGB}{255, 183, 77}
\definecolor{orangeIV}{RGB}{255, 167, 38}
\definecolor{orangeV}{RGB}{255, 152, 0}
\definecolor{orangeVI}{RGB}{251, 140, 0}
\definecolor{orangeVII}{RGB}{245, 124, 0}
\definecolor{orangeVIII}{RGB}{239, 108, 0}
\definecolor{orangeIX}{RGB}{230, 81, 0}
\definecolor{orangeAI}{RGB}{255, 209, 128}
\definecolor{orangeAII}{RGB}{255, 171, 64}
\definecolor{orangeAIV}{RGB}{255, 145, 0}
\definecolor{orangeAVII}{RGB}{255, 109, 0}

\definecolor{deeporangeOV}{RGB}{251, 233, 231}
\definecolor{deeporangeI}{RGB}{255, 204, 188}
\definecolor{deeporangeII}{RGB}{255, 171, 145}
\definecolor{deeporangeIII}{RGB}{255, 138, 101}
\definecolor{deeporangeIV}{RGB}{255, 112, 67}
\definecolor{deeporangeV}{RGB}{255, 87, 34}
\definecolor{deeporangeVI}{RGB}{244, 81, 30}
\definecolor{deeporangeVII}{RGB}{230, 74, 25}
\definecolor{deeporangeVIII}{RGB}{216, 67, 21}
\definecolor{deeporangeIX}{RGB}{191, 54, 12}
\definecolor{deeporangeAI}{RGB}{255, 158, 128}
\definecolor{deeporangeAII}{RGB}{255, 110, 64}
\definecolor{deeporangeAIV}{RGB}{255, 61, 0}
\definecolor{deeporangeAVII}{RGB}{221, 44, 0}

\definecolor{brownOV}{RGB}{239, 235, 233}
\definecolor{brownI}{RGB}{215, 204, 200}
\definecolor{brownII}{RGB}{188, 170, 164}
\definecolor{brownIII}{RGB}{161, 136, 127}
\definecolor{brownIV}{RGB}{141, 110, 99}
\definecolor{brownV}{RGB}{121, 85, 72}
\definecolor{brownVI}{RGB}{109, 76, 65}
\definecolor{brownVII}{RGB}{93, 64, 55}
\definecolor{brownVIII}{RGB}{78, 52, 46}
\definecolor{brownIX}{RGB}{62, 39, 35}

\definecolor{grayOV}{RGB}{250, 250, 250}
\definecolor{grayI}{RGB}{245, 245, 245}
\definecolor{grayII}{RGB}{238, 238, 238}
\definecolor{grayIII}{RGB}{224, 224, 224}
\definecolor{grayIV}{RGB}{189, 189, 189}
\definecolor{grayV}{RGB}{158, 158, 158}
\definecolor{grayVI}{RGB}{117, 117, 117}
\definecolor{grayVII}{RGB}{97, 97, 97}
\definecolor{grayVIII}{RGB}{66, 66, 66}
\definecolor{grayIX}{RGB}{33, 33, 33}

\definecolor{bluegrayOV}{RGB}{236, 239, 241}
\definecolor{bluegrayI}{RGB}{207, 216, 220}
\definecolor{bluegrayII}{RGB}{176, 190, 197}
\definecolor{bluegrayIII}{RGB}{144, 164, 174}
\definecolor{bluegrayIV}{RGB}{120, 144, 156}
\definecolor{bluegrayV}{RGB}{96, 125, 139}
\definecolor{bluegrayVI}{RGB}{84, 110, 122}
\definecolor{bluegrayVII}{RGB}{69, 90, 100}
\definecolor{bluegrayVIII}{RGB}{55, 71, 79}
\definecolor{bluegrayIX}{RGB}{38, 50, 56}
\definecolor{bluegrayX}{RGB}{17, 23, 26}

\definecolor{myACMBlue}{cmyk}{1,0.1,0,0.1}
\definecolor{myACMYellow}{cmyk}{0,0.16,1,0}
\definecolor{myACMOrange}{cmyk}{0,0.42,1,0.01}
\definecolor{myACMRed}{cmyk}{0,0.90,0.86,0}
\definecolor{myACMLightBlue}{cmyk}{0.49,0.01,0,0}
\definecolor{myACMGreen}{cmyk}{0.20,0,1,0.19}
\definecolor{myACMPurple}{cmyk}{0.55,1,0,0.15}
\definecolor{myACMDarkBlue}{cmyk}{1,0.58,0,0.21}
\definecolor{myReferenceURLColor}{HTML}{09326F}

\hypersetup{
  colorlinks,
  linkcolor=blueVIII,
  citecolor=blueVIII,
  urlcolor=myReferenceURLColor,
  filecolor=myACMDarkBlue
}

\newcommand{\todonow}[1]{\textcolor{pinkIV}{[#1]}}
\newcommand{\todo}[1]{\textcolor{pinkIV}{[#1]}}
\newcommand{\tocite}[1]{\textcolor{greenVI}{[Cite #1]}}

\newcommand{\seongmin}[1]{\textcolor{orangeVI}{[#1 -seongmin]}}
\newcommand{\david}[1]{\textcolor{blueVI}{[#1 -david]}}
\newcommand{\polo}[1]{\textcolor{tealV}{[#1 -polo]}}

\newcommand{\mkclean}{
    \renewcommand{\todonow}[1]{}
    \renewcommand{\todo}[1]{}
    \renewcommand{\tocite}[1]{}
  \renewcommand{\david}[1]{}
  \renewcommand{\polo}[1]{}
  \renewcommand{\seongmin}[1]{}
}

\mkclean

\newcommand{\tool}{\textsc{Mobile Fitting Room}} 
\begin{document}

\title{\tool{}: On-device Virtual Try-on via Diffusion Models}
\settopmatter{authorsperrow=4}

\author{Justin Blalock}
\email{jblalock30@gatech.edu}
\affiliation{%
  \institution{Georgia Tech}
  \city{Atlanta}
  \state{Georgia}
  \country{USA}
}

\author{David Munechika}
\email{david.munechika@gatech.edu}
\affiliation{%
  \institution{Georgia Tech}
  \city{Atlanta}
  \state{Georgia}
  \country{USA}
}

\author{Harsha Karanth}
\email{hkaranth3@gatech.edu}
\affiliation{%
  \institution{Georgia Tech}
  \city{Atlanta}
  \state{Georgia}
  \country{USA}
}

\author{Alec Helbling}
\email{alechelbling@gatech.edu}
\affiliation{%
  \institution{Georgia Tech}
  \city{Atlanta}
  \state{Georgia}
  \country{USA}
}

\author{Pratham Mehta}
\email{pratham@gatech.edu}
\affiliation{%
  \institution{Georgia Tech}
  \city{Atlanta}
  \state{Georgia}
  \country{USA}
}
  
\author{Seongmin Lee}
\email{seongmin@gatech.edu}
\affiliation{%
  \institution{Georgia Tech}
  \city{Atlanta}
  \state{Georgia}
  \country{USA}
}

\author{Duen Horng Chau}
\email{polo@gatech.edu}
\affiliation{%
  \institution{Georgia Tech}
  \city{Atlanta}
  \state{Georgia}
  \country{USA}
}

\renewcommand{\shortauthors}{Blalock, et al.}

\begin{abstract}
The growing digital landscape of fashion e-commerce calls for interactive and user-friendly interfaces for virtually trying on clothes. Traditional try-on methods grapple with challenges in adapting to diverse backgrounds, poses, and subjects. While newer methods, utilizing the recent advances of diffusion models, have achieved higher-quality image generation,  the human-centered dimensions of mobile interface delivery and privacy concerns remain largely unexplored. We present \tool{}, the first on-device diffusion-based virtual try-on system. To address multiple inter-related technical challenges such as high-quality garment placement and model compression for mobile devices, we present a novel technical pipeline and an interface design that enables privacy preservation and user customization. A usage scenario highlights how our tool can provide a seamless, interactive virtual try-on experience for customers and provide a valuable service for fashion e-commerce businesses.
\end{abstract}



\begin{teaserfigure}
    \centering
    \includegraphics[width=0.8\linewidth]{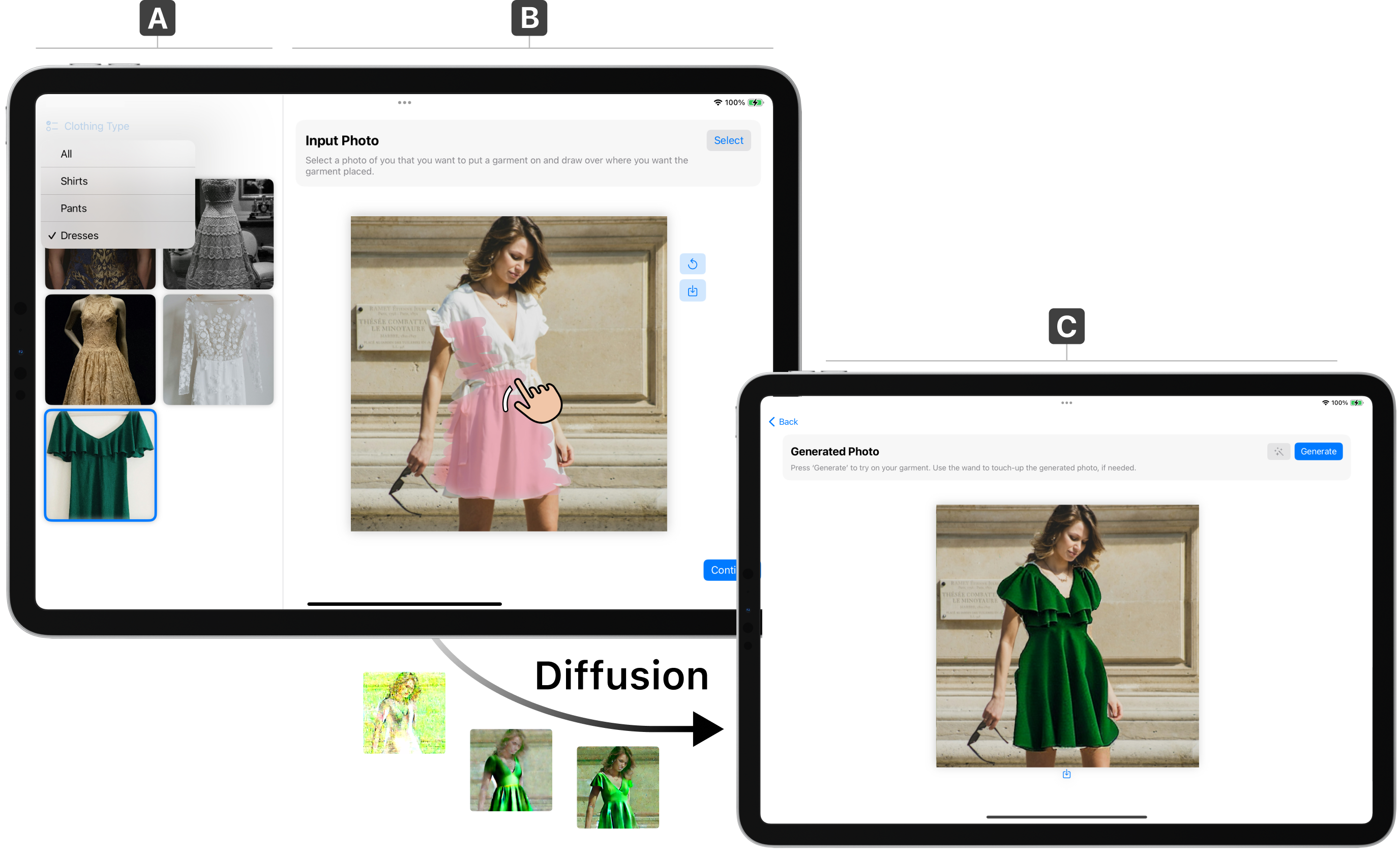}
    \caption{With a novel approach to virtual try-on experiences, \tool{} is a mobile app that overcomes the formidable challenges of on-device machine learning in order to empower privacy-preserving and personalized virtual try-on. Consider the task of trying on a dress before buying it. 
    \textbf{(A)}\label{fig:app-large-A} In the \textit{Garments Window}, users can filter through clothes in the catalog and select the dress that they wish to try on. 
    \textbf{(B)} After the user selects the garment, the \textit{Personalization Window} allows them to control and customize where the dress is placed by using their finger to draw over the desired area. \textbf{(C)} The user then navigates to the  \textit{Result Page} to view the generated try-on images.}
    \label{fig:app-large}
    \vspace{8pt}
\end{teaserfigure}

\maketitle
\pagestyle{plain}

\section{Introduction}

As the fashion world shifts towards the digitization of shopping experiences and the rapid growth of the fashion e-commerce industry \cite{statista}, the need for interactive and user-friendly interfaces for trying on clothes becomes increasingly paramount. Virtual try-on aims to mimic the in-store fitting experience by showing the user an image of how they would look wearing a specific clothing item.

Research efforts that have explored virtual try-on traditionally use warping networks \cite{choi2021viton, raj2018swapnet, ge2021parser} and other methods that learn to geometrically transform the input garment to align with the subject. 
These warping modules often struggle to cope with non-rigid transformations, such as complex body poses \cite{li2023warpdiffusion}. 
On the other hand, diffusion models have gained significant popularity for their ability to generate creative, high-quality images \cite{huang2023composer, qin2023unicontrol, ramesh2022hierarchical, rombach2022high, saharia2022photorealistic, zhang2023adding}. 
Downstream applications of diffusion models have aimed to further improve the user's control over the model's output. Some examples include image editing \cite{hertz2022prompt, tumanyan2023plug}, inpainting \cite{lugmayr2022repaint, xie2023smartbrush, yu2023inpaint}, and controlled generation \cite{qin2023unicontrol, mou2023t2i}. 
With these methods, researchers have successfully extended diffusion models to support virtual try-on \cite{choi2021viton, zheng2019virtually, li2023warpdiffusion, zhu2023tryondiffusion, kim2023stableviton}.
However, many of these studies do not consider the human-centered component of try-on, such as how to deliver this experience to the end user.
With a significant portion of shopping occurring on users' mobile devices \cite{devicestat}, it becomes crucial to explore how to best deliver the try-on experience on a mobile interface. Furthermore, since try-on involves private user information (their pictures), careful consideration must also be taken to preserve privacy.

We argue that running try-on machine learning (ML) models on-device is a significant step towards improving user privacy and experience. 
On-device ML refers to running models directly on a user's (portable) device such as a tablet or phone. 
There are several reasons why running try-on models on-device benefits users. First, running ML models in the cloud presents a privacy risk to the user. 
The user's images must leave their device to be stored or processed in a cloud server, which may lack integrity \cite{huRui}. 
When running models on-device, sensitive information never has to leave the user's device.
Second, on-device ML models result in gains for offline application use, eliminating the need for an internet connection to upload images and download server results \cite{hohman2023model, dhar2021survey}.

However, running these models on-device presents a significant challenge since modern-day ML models often have hundreds of millions to tens of billions of parameters \cite{villalobos2022machine} which consume large amounts of device resources \cite{appleOptimize, jeong2018communication};
the model used in this paper is no exception. It has approximately 1.2 billion parameters \cite{stable-diffusion-coreml-apple-silicon}.
Therefore, in this work, we employ techniques such as palettization \cite{hohman2023model, grønlund2018fast} and chunking to reduce the try-on ML model size by 36\% with a minimal drop in accuracy. Additionally, we use other techniques to speed up inference for mobile devices.

Our work bridges the gap between diffusion-based try-on efforts \cite{ li2023warpdiffusion, zhu2023tryondiffusion, kim2023stableviton} and pragmatic mobile experiences. In this ongoing work, we contribute:

\begin{enumerate}
    \item \textbf{A fully on-device deployment of a diffusion-based model for virtual try-on} that enables novel virtual try-on experiences. 
    We present \tool{} (Figure~\ref{fig:app-large}), an iPadOS app that exemplifies how these experiences can be bolstered by the interaction afforded to portable devices (Section \ref{user-interface}). 
To the best of our knowledge, this is the first research effort to synthesize mobile diffusion with personalized, controlled generation for the purpose of virtual try-on applications. 
    \item \textbf{A novel technical pipeline critical for enabling efficient on-device try-on models} (Figure \ref{fig:pipeline}). This pipeline paves the way for on-device virtual try-on by tackling several challenges including fine-tuning a diffusion model to allow us to control which garment is placed on a person (Section \ref{fine-tune}), compressing the model to retain quality while meeting the storage and efficiency demands of mobile devices (Section \ref{conversion}), and combining conditional control with novel user interaction to enable user personalization of the virtual try-on experience (Section \ref{inpainting}).
\end{enumerate}

\section{User Interface} \label{user-interface}

\subsection{Overview}

We present \tool{} as a mobile application that allows users to see how they would look when wearing various garments. The goal of our application is to allow users to try-on clothes in an interactive manner that also preserves user privacy. A typical user flow is to select a garment, select a picture of oneself, draw on the picture to indicate which specific area should be replaced with the chosen garment, and click ``Generate.'' At this point, the application will generate an image of the chosen garment on the user. The demo mobile application is implemented using Swift and is built for iPadOS. The application has three components:
\begin{enumerate}
    \item \textbf{Garments Window} shows the list of garments that a user can choose to try on.
    \item \textbf{Personalization Window} allows the user to select an image of themself and draw over the region that they want replaced.
    \item \textbf{Result Page} shows the generated try-on image and allows the user to make any necessary touch-ups.
\end{enumerate}

\subsection{Garments Window}

The \textit{Garments Window} (Figure \ref{fig:app-large}A) provides users with a simple way to select a garment to try-on. Garments that have a corresponding try-on ML model (Section \ref{fine-tune}) are populated into the app. A fashion e-commerce store, for instance, could upload their clothing catalog to the app for users to try on. Users can use the dropdown menu to limit their search to clothing of only one type.

\subsection{Personalization Window}

\begin{figure}[t]
    \centering
    \includegraphics[width=1\linewidth]{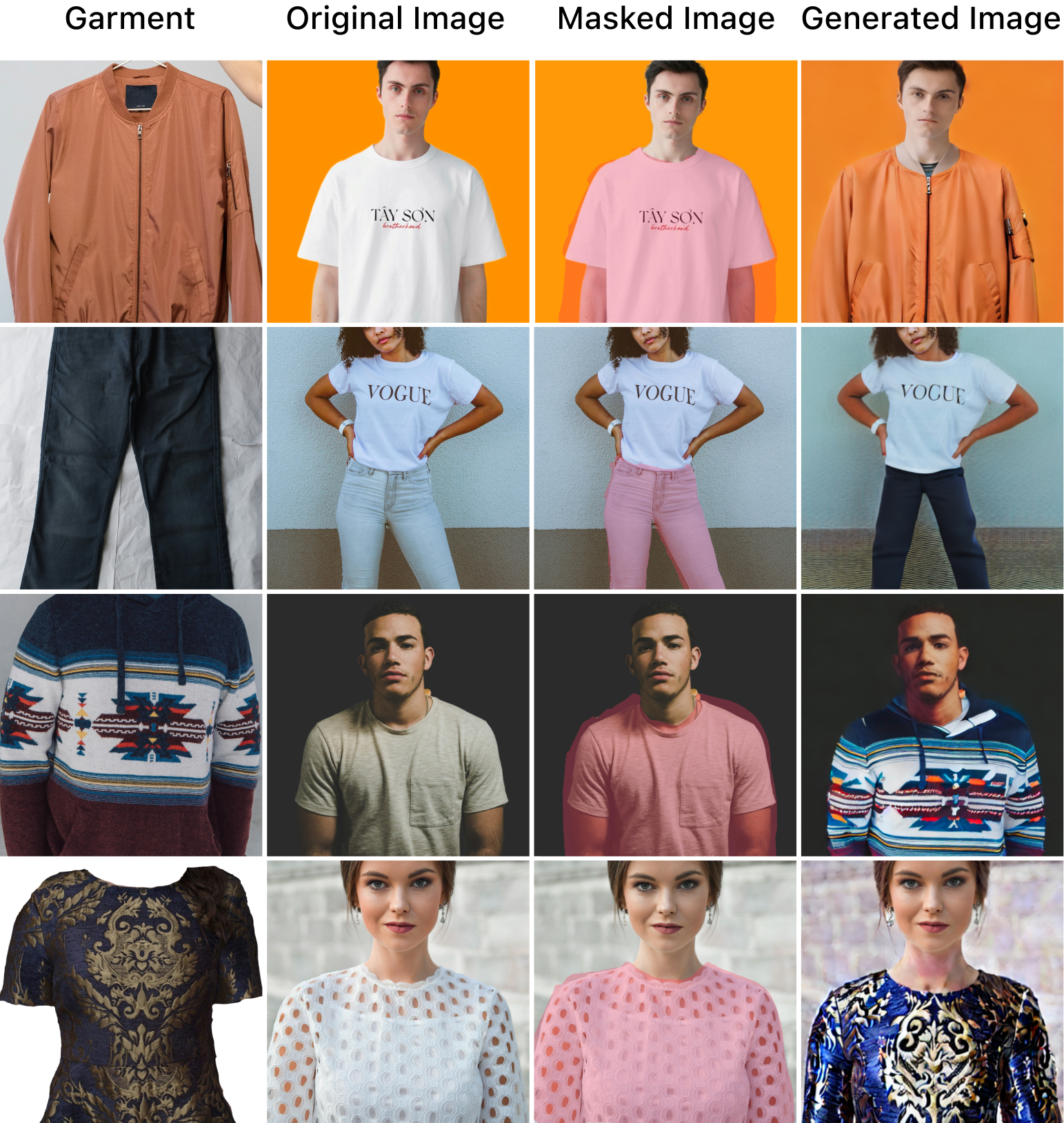}
    \caption{Samples of generated images from \tool{}. The user inputs an image and draws a mask around the area that they want replaced.
    }
    \label{fig:results}
\end{figure}

The \textit{Personalization Window} (Figure \ref{fig:app-large}B) allows users to input a photo of themselves and mask out the garment that they want changed. Users can do this by selecting a photo from their camera roll and then use their finger or a stylus to draw where they want the new garment to be placed. For example, if they wanted to replace the shirt that they are wearing in their input photo, they would simply draw over that shirt (Figure~\ref{fig:results}). By drawing over the area that they want replaced, users are able to use a unique form of interaction to add personalization to the output image. They may want to see what the garment would look like when arranged a specific way or they may want to see what creative fittings the diffusion process can produce. With this interaction, users are given an additional degree of control and creativity not found in current virtual try-on endeavors.

\subsection{Result Page}

The \textit{Result Page} (Figure \ref{fig:app-large}C) shows the generated try-on image. Users can use the eraser tool to clean up any artifacts left from the diffusion process. We have found that if the mask that the user draws over themselves is too obscure, then the inpainting process may cause irregularities such as uneven transitions from the masked to the non-masked area. Further, if the try-on ML model has been compressed to an extreme degree, the inpainting process sometimes slightly alters faces and the hue of the background.  Therefore post-processing via the eraser tool may need to be done in order to perfect the image. We use the same drawing interaction to allow the user to make these touch-ups. By drawing over the generated image, the generated image is blended with the original image in order to ``restore'' elements that were modified through diffusion. Additionally, this ability to restore elements of the original photo provides a counterbalance to the user's freedom to choose their own mask. In the event that they have made an ineffective mask, instead of regenerating a new image, they can simply remove blemishes of the current generated image.

\section{Techniques for On-Device Controlled Diffusion}
Running the try-on diffusion model on-device presents an opportunity for preserving privacy and removing the reliance on internet connectivity \cite{hohman2023model, dhar2021survey}. Though, diffusion models are large and computationally expensive, thus presenting many challenges to get them to run on mobile devices \cite{hohman2023model}. The challenges that we face are: how do we train an ML model to place specific garments on people (Section \ref{fine-tune}), how do we convert the trained model to a usable form on mobile devices (Section \ref{conversion}), how do we compress the model so that it uses less resources (Section \ref{conversion}), and how do we use the on-device model to draw garments on people in a time efficient manner (Section \ref{inpainting}).

\begin{figure*}[t]
    \includegraphics[width=0.8\linewidth]{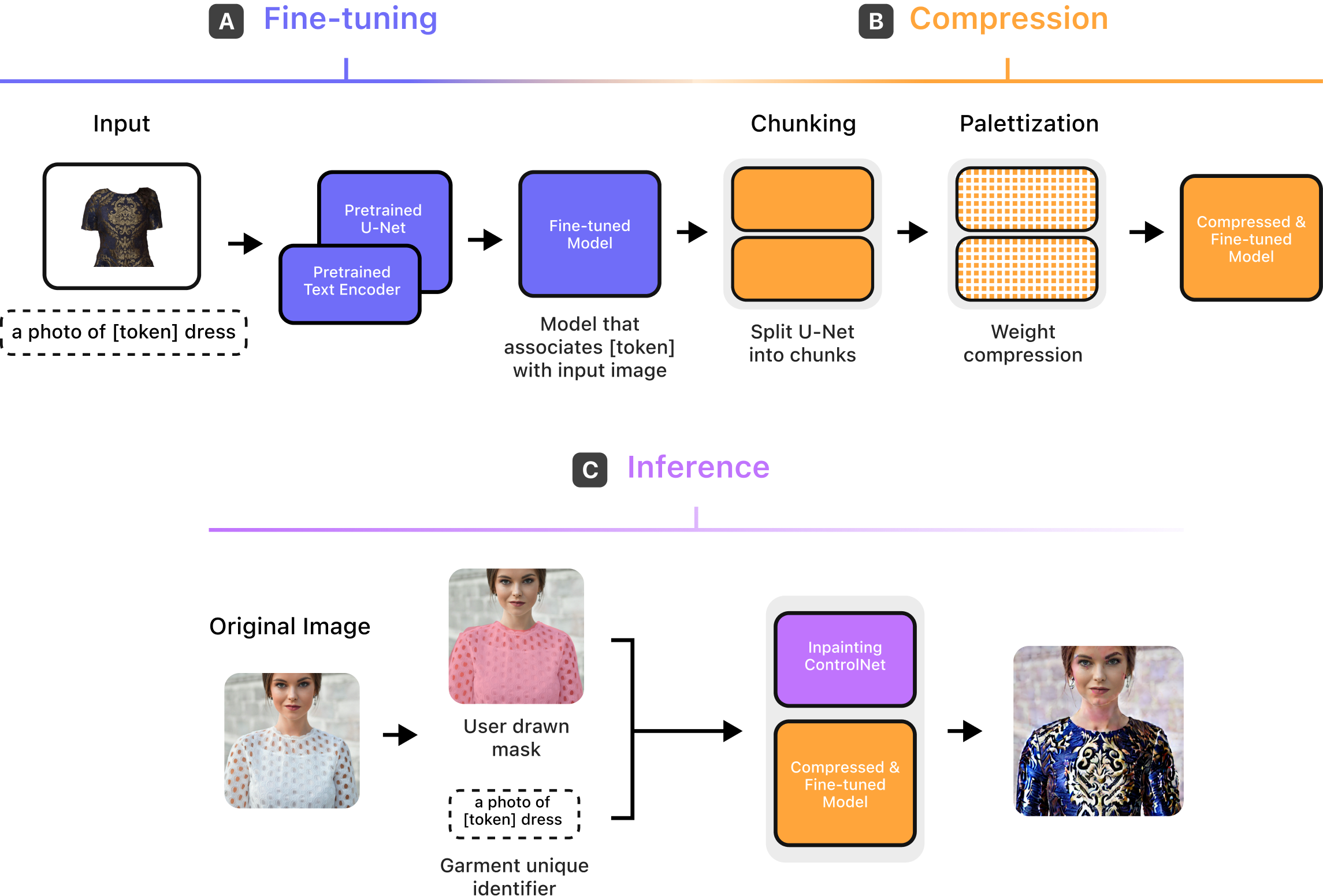}
    \caption{The pipeline for \tool{} allows for an on-device privacy-preserving model whose flexibility allows for user personalization through interaction. (A) Fine-tuning generates a model that allows us to summon a specific garment with a unique identifier. (B) We then compress the model so that it can meet the storage and efficiency demands of a mobile device. (C) \tool{} allows users to specify where they want a garment to be placed by using ControlNet inpainting. }
    \label{fig:pipeline}
\end{figure*}

\subsection{Model Fine-Tuning} \label{fine-tune}

The first challenge is to create an ML model capable of generating specific clothing items. To achieve this, we utilize fine-tuning, a method to further train an ML model to obtain specialized outputs (Figure~\ref{fig:pipeline}A). We chose to fine-tune Stable Diffusion v1-5 \cite{rombach2022high}, a pretrained text-to-image diffusion model. Text-to-image diffusion models learn to iteratively denoise randomly sampled inputs conditioned on a text prompt. We utilize Dreambooth \cite{ruiz2022dreambooth}, a method to fine-tune diffusion models, to add reference images into new environments, also known as subject-driven generation. This method involves fine-tuning the U-Net --- the main neural network component of the diffusion process pivotal in image generation  --- on a small number of reference images and fine-tuning the text encoder on text prompts that uniquely identify the reference object. The text prompts contain the identifier for the reference object plus the class that the reference object belongs to. In our case, the reference object is a garment and the text prompt that we fine-tune with could be "a photo of an \textit{rtr} shirt" where \textit{rtr} is the identifier. By doing this, we can evoke a specific garment when a unique identifier is used.

One limitation of this approach is that it is prone to overfitting, especially in the case of try-on where at inference time the prompts are similar to the contexts in which they were trained \cite{ruiz2022dreambooth}. One way that we found to mitigate artifacts from overfitting is to use only images of garments themselves as inputs to fine-tuning, not images of people wearing the garments. An additional technique that we used was to decrease the guidance weight during the diffusion process \cite{ho2022classifier}. 

Another limitation that we discovered through our testing is that the DreamBooth \cite{ruiz2022dreambooth} technique did not work well for multiple subjects. In other words, training on multiple garments produced low-quality results, likely due to catastrophic forgetting of prior garments where each garment overwrites previous ones in the model's ``memory'' \cite{kirkpatrick2017overcoming}. 
Therefore, the mobile app incorporates a model for each clothing item. Utilizing the compression techniques outlined in Section \ref{conversion}, this approach works well for managing small catalogs of garments.

\subsection{Model Conversion and Inference}\label{conversion}

After obtaining the fine-tuned diffusion model, it is not currently in a form usable by iPadOS and it also would be too memory, storage, and compute intensive to run on-device. Therefore, we apply several model compression techniques to make the model more efficient.

We utilize the open-source CoreML Stable Diffusion project \cite{stable-diffusion-coreml-apple-silicon} to compress and transform the model (Figure~\ref{fig:pipeline}B). First, we cluster (palettize) the weights of the U-Net and text encoder using k-means \cite{grønlund2018fast} which compresses the model \cite{hohman2023model, cho2022dkm}. In our testing, we were able to reduce model size by 36\% (and retain image quality) by using 6-bit palettization. Secondly, since the U-Net is a very large tensor, we split the U-net into two similarly-sized chunks, thus increasing L2 cache residency and multi-core utilization \cite{appleMLR}. Third, we employ the split einsum attention implementation which reduces the number of memory copies during computation and increases the efficiency of tensor operations, resulting in speed and memory improvements for mobile devices
\cite{stable-diffusion-coreml-apple-silicon, appleMLR, klaus2023compiling}.

At this point, we have reduced the model size and increased computational efficiency significantly. The next step is to compile the model and add it to the app's document directory. As noted in Section \ref{fine-tune}, each garment is currently associated with one model. In order to provide the best user experience, several strategies can be employed here to balance app size and download time. To reduce app size, we could only download models for specific clothing items on demand. This way, the app can be shipped with a small size. The drawback of this approach is that downloading a model in-app may take many seconds or minutes depending on the model size. A compromise to this is to identify the garments that the user is most interested in and download this subset in the background. This approach strikes a balance between app size and download wait time.

\subsection{Inpainting} \label{inpainting}

Since we have a model that can conjure specific garments at will, we need a mechanism for the user to add their own personalized touch and to control the image generation process. We found that allowing the user to draw the specific region that they wanted to be replaced allows for a simple and easily controllable way to achieve this goal (Figure~\ref{fig:pipeline}C). In Section \ref{planned-eval} we discuss our planned usability study for this feature.

To replace a specified region of the input image with a new piece of clothing, we must add conditional control to the model \cite{zhang2023adding}. We do so by using an inpainting model; an inpainting model is capable of ``filling in'' masked out portions of an image. Inpainting signals to the generation that we only want to alter the subset of pixels pertaining to the mask. This behavior is desired since we want to preserve everything in the image except the region that has been masked out. In sum, we use the unique identifier corresponding to the garment (Section \ref{fine-tune}) which informs the diffusion model to produce the garment and the inpainting module ensures that this garment is placed in the location specified by the user.

\section{Preliminary Result: Virtual Try-On Usage Scenario for Retail Business}
We present a dual-perspective usage scenario of deploying an app that uses \tool{}: first from the perspective of the business deploying an app with virtual try-on, and next from the viewpoint of the customer using the app.
The business perspective highlights how \tool{} provides an inexpensive way to add user value to the online shopping experience. 
The user perspective illustrates the benefits that users gain from the privacy-preserving and interactive features of \tool{}.

Bloom is a small business that has a retail store and a mobile app. 
Bloom has been experiencing a sudden increase of online visitors to their app but a decrease in people coming to the store in-person. 
As such, the customer experience has been negatively impacted since customers cannot try-on clothes, resulting in an increase in refunds. 
Further, in order to compete with other fashion brands, Bloom needs an inexpensive way to improve online customer satisfaction and drive growth. 
In order to promote a similar experience of trying on clothes in-store through their app, Bloom uses the technology from \tool{} to enable customers to virtually try-on clothes. 
First, the employees at Bloom select the most popular garments from their catalog for users to virtually try-on. 
Then, their technology specialist uses the methods described in Section \ref{fine-tune} to create specialized models for the garments. 
The specialist compresses and converts the models using the techniques described in Section \ref{conversion} since the models were originally too large and computationally expensive to run on device. They then incorporate the interface implementation of \tool{} into their existing app in order to add interactivity and customization to the fitting experience.
Following the addition of this fitting experience, online customer satisfaction is at an all-time high, and their online store has had exponential growth due to the novelty of this feature. Also, since the models run on-device, Bloom enjoys higher profit margins since they need not pay for servers.

When Rachel, a customer, opens the Bloom app, she navigates through the dresses in the \textit{Garments Window} (Figure \ref{fig:app-large}A). She sees a dress that she likes, but she is hesitant to buy it—since she needs the dress for a dance, it is essential that it fits and looks good on her. 
After selecting the dress, Rachel opens the \textit{Personalization Window} (Figure \ref{fig:app-large}B) and uploads a picture of herself. 
She likes to wear dresses in a unique way as a form of self-expression. Thus, she draws over the uploaded image in the way that she envisions wearing the dress.
Rachel then navigates to the \textit{Result Page} (Figure \ref{fig:app-large}C) and generates an image of herself wearing the dress. 
Before saving the image, she uses the eraser tool to polish it. 
Rachel sees that the dress fits well and also has the potential to be styled in her own custom way, giving her confidence in buying the dress despite not being able to try it on in-person. 
She also appreciates that her photos were not sent to an external server as she is highly conscious of her privacy and would not have wanted to upload her private photos to a unknown server. Rachel is now able to buy the dress with peace of mind.

\section{Ongoing Work}

\subsection{Model Improvements}

Using model compression techniques to reduce the size of the diffusion model can affect the quality of the generated image especially in our case where we first fine-tune on only a few images and then we compress. Additionally, we have found that low-bit quantization or palettization sometimes produces images with altered hues or uncanny faces. We are currently exploring a various number of pipelines to address this issue and improve overall model fidelity. For example, recent works have shown success in fine-tuning language models after they were already compressed \cite{liu2021enabling} and we hypothesize a similar approach may be promising for diffusion models as well. In the same way that fine-tuning and pruning are used in conjunction when efficiently compressing neural networks \cite{martinez2021permute, tung2017fine}, maintaining the diffusion model fidelity may require joint compression and fine-tuning in the process.

Furthermore, the current version of \tool{} requires fine-tuning a separate diffusion model for each garment. This does not scale well for large garment catalogs and also means that users cannot upload their own garments. To rectify this, we plan to improve our system to support zero-shot generation so that users can virtually try-on clothing items that the diffusion model has not been fine-tuned on; additionally, the model will also only need to be fine-tuned once. Recent research \cite{chen2023anydoor} has seen success in incorporating identity and detail features to support zero-shot generation that, after training on a collection of try-on data, generalizes well to various gestures, colors, and patterns. We plan to try a similar approach but intend to retain our element of personalization by allowing the user to customize how the garment fits. Regardless of the pipeline, users should be able to interact with the system to reinforce a creative experience.

\subsection{Planned Evaluation}\label{planned-eval}

We plan to conduct a user study to evaluate the usability and usefulness of \tool{}. The study will be divided into two stages.

In the first stage, we will recruit participants from our local community who engage in fashion e-commerce. We will ask participants to complete a set of tasks. One task, for instance, will ask participants to generate a photo of themself with a different shirt; to do this, they will need to navigate the interface in order to select a shirt from the \textit{Garments View}, then select a picture of themself and draw a mask in the \textit{Personalization View}, and generate an image on the \textit{Result Page}. From our simple set of tasks, we will be able to collect both qualitative (e.g., satisfaction, learnability) and quantitative data (e.g., the amount of time it takes to complete tasks, operations used) from our observations. This will be a small-scale, in-person study allowing us to easily observe user behavior and facilitating detailed data collection and reducing recall bias. 

In the second stage of the study, we will recruit a large number of participants from the Shopify community and from Facebook fashion e-commerce groups.
We will upload the mobile app to Testflight \cite{testflight} and collect user analytics data. This will enable us to easily distribute the app to a large amount of testers and track user interaction patterns and behaviors that are revealed through analytics. %

At the end of each stage, we will provide a survey to participants and ask them to (1) evaluate the usability of our app and (2) suggest improvements to existing features or additions of new features.

\subsection{Deployment}

To broaden impact, we plan to deploy \tool{} as a mobile application on the App Store and make it usable for all iPadOS users. Users will be able to use our app to virtually try-on various clothing items completely on device without sending their personal pictures to any external server. We also intend to open-source the code for \tool{}.

\section{Conclusion}

We present \tool{}, an on-device app for virtual try-on powered by diffusion models. We have plans to improve \tool{} to increase its usability for a wide range of subjects by supporting zero-shot generation along with higher generation quality. We will evaluate \tool{} through a user study and deploy it as a mobile app on the App Store.

\bibliographystyle{ACM-Reference-Format}
\bibliography{apple-ai}

\end{document}